\documentclass[aps,pra,nofootinbib]{revtex4}

\usepackage{amsmath}
\usepackage{graphicx}

\newcommand{\be}{\begin{equation}}
\newcommand{\ee}{\end{equation}}
\newcommand{\bearray}{\begin{eqnarray}}
\newcommand{\eearray}{\end{eqnarray}}
\newcommand{\bse}{\begin{subequations}}
\newcommand{\ese}{\end{subequations}}
\newcommand{\bcases}{\begin{cases}}
\newcommand{\ecases}{\end{cases}}

\begin{document}

\title{Angular decomposition of tensor products of a vector}

\author{Gregory S. Adkins}
\affiliation{Franklin \& Marshall College, Lancaster, Pennsylvania 17604 USA}
\email[]{gadkins@fandm.edu}
\date{\today}

\begin{abstract}
The tensor product of $L$ copies of a single vector, such as $p_{i_1} \cdots p_{i_L}$, can be analyzed in terms of angular momentum.  When $p_{i_1} \cdots p_{i_L}$ is decomposed into a sum of components $\left ( p_{i_1} \cdots p_{i_L} \right )^L_\ell$, each characterized by angular momentum $\ell$, the components are in general complicated functions of the $p_i$ vectors, especially so for large $\ell$.  We obtain a compact expression for $\left ( p_{i_1} \cdots p_{i_L} \right )^L_\ell$ explicitly in terms of the $p_i$ valid for all $L$ and $\ell$.  We use this decomposition to perform three-dimensional Fourier transforms of functions like $p^n \hat p_{i_1} \cdots \hat p_{i_L}$ that are useful in describing particle interactions.
\end{abstract}


\maketitle


\section{Introduction}
\label{introduction}

Three-dimensional Fourier transforms of the general form
\be \label{FTC}
I_{n; i_1 \cdots i_L}(\vec r\, ) = \int \frac{d^3 p}{(2 \pi)^3} e^{i \vec p \cdot \vec r} \, p^n \hat p_{i_1} \cdots \hat p_{i_L}
\ee
(where $\hat p = \vec p/p$) have a wide variety of uses.  For example,
\be \label{FTell0}
I_{-2}(\vec r\, ) = \int \frac{d^3 p}{(2 \pi)^3} e^{i \vec p \cdot \vec r} \, \frac{1}{p^2} = \frac{1}{4 \pi r}
\ee
is the Fourier representation of the Coulomb potential.  Two related transforms that occur in the study of fermion-fermion interactions \cite{Breit29,Bethe57,Berestetskii82} are
\bse \label{FT}
\bearray
I_{-1; i}(\vec r\, ) &=& \int \frac{d^3 p}{(2 \pi)^3} e^{i \vec p \cdot \vec r} \, \frac{p_i}{p^2} = \frac{i \hat x_i}{4 \pi r^2} , \label{FTell1} \\
I_{0; i j}(\vec r\, ) &=& \int \frac{d^3 p}{(2 \pi)^3} e^{i \vec p \cdot \vec r} \, \frac{p_i p_j}{p^2} = \frac{1}{3} \delta_{i j} \delta(\vec r \,) - \frac{3}{4 \pi r^3} \left ( \hat x_i \hat x_j - \frac{1}{3} \delta_{i j} \right ) . \label{FTell2} 
\eearray
\ese
(In our notation the position vector has components  $\vec r = (x_1,x_2,x_3)=(x, y, z)$, and $\hat r = \vec r/r$ with components $\hat x_i = x_i/r$ is the associated unit vector.)  The structure of three dimensional Fourier transforms such as \eqref{FTC} is organized by angular momentum.  One sees that both the original function $1/p^2$ and the transform $1/(4 \pi r)$ of \eqref{FTell0} are scalars under rotation.  The original function $p_i/p^2$ of \eqref{FTell1} is a vector with $\ell=1$ because the components $\hat p_i$ can be expressed linearly in terms of spherical harmonics $Y_1^m(\hat p)$ with $\ell=1$.  The transform $i \hat x_i/(4 \pi r^2)$ of \eqref{FTell1} also has $\ell=1$ as $\hat x_i$ can be expressed linearly in terms of $Y_1^m(\hat r)$.  The original function in \eqref{FTell2}, $\hat p_i \hat p_j$, is a combination of $\ell=0$ and $\ell=2$:
\be \label{decompell2}
\hat p_i \hat p_j = \left ( \hat p_i \hat p_j \right )^2_0 + \left ( \hat p_i \hat p_j \right )^2_2 = \left ( \frac{1}{3} \delta_{i j} \right ) + \left ( \hat p_i \hat p_j - \frac{1}{3} \delta_{i j} \right )
\ee
where $\left ( \hat p_i \hat p_j \right )^2_0 = \frac{1}{3} \delta_{i j}$ is the $\ell=0$ component and $\left ( \hat p_i \hat p_j \right )^2_2 = \hat p_i \hat p_j - \frac{1}{3} \delta_{i j}$ is the $\ell=2$ component.  (In general we will write $\left ( p_{i_1} \cdots p_{i_L} \right )^L_\ell$ for the component of $p_{i_1} \cdots p_{i_L}$ of angular momentum $\ell$.)  We know that $\left ( \hat p_i \hat p_j \right )^2_2$ has $\ell=2$ because it can be expressed linearly in terms of $Y_2^m(\hat p)$: $\left ( \hat p_i \hat p_j \right )^2_2 = \sum_m C^{2 m}_{i j} Y_2^m(\hat p)$.  It is apparent that the $\ell=0$ and $\ell=2$ components of $\hat p_i \hat p_j$ behaves differently under the Fourier transform, acquiring different radial factors.  It is generally true that in transforms like \eqref{FTC} it is useful to decompose $\hat p_{i_1} \cdots \hat p_{i_L}$ into components of definite $\ell$ and deal with each component separately.

The purpose of this work is to show how the decomposition of $p_{i_1} \cdots p_{i_L}$ can be done and to give explicit expressions for the components of $\left ( p_{i_1} \cdots p_{i_L} \right )^L_\ell$ having various values of angular momentum $\ell$.  This is a generalization of \eqref{decompell2} to an arbitrary number of vectors $L$.  Our derivations are presented in terms of unit vectors because the relation
\be
\left ( p_{i_1} \cdots p_{i_L} \right )^L_\ell = p^L \left ( \hat p_{i_1} \cdots \hat p_{i_L} \right )^L_\ell 
\ee
allows us to immediately obtain the general case.

This work is organized as follows.  In \ref{sec2} we will review the method for performing three-dimensional Fourier transforms like \eqref{FTC} making use of angular decomposition.  In \ref{sec3} we obtain the general expression for the component of $\hat p_{i_1} \cdots \hat p_{i_L}$ of angular momentum $\ell$.  Finally, in \ref{sec4}, we give some examples and applications of our results.

More general studies of the relation between Cartesian and spherical components of tensors have been done, \cite{Coope65,Coope70,Stone75,Stone76,Normand82} but the results of those studies are not in a form useful for our purposes here.


\section{Three-dimensional Fourier transforms using angular decomposition}
\label{sec2}

A systematic procedure exists for the evaluation of transforms such as \eqref{FTC} based on the decomposition of $\hat p_{i_1} \cdots \hat p_{i_L}$ into components of definite angular momentum. \cite{Adkins13} Our purpose in this section is to review this procedure.  We begin by noting that any function of angles, such as $\hat p_{i_1} \cdots \hat p_{i_L}$, can be written in terms of spherical harmonics:
\be \label{psintermsofY}
\hat p_{i_1} \cdots \hat p_{i_L} = \sum_{\ell=L}^{1 \, {\rm or} \, 0} \sum_{m=-\ell}^\ell C^{\ell m}_{i_1 \cdots i_L} Y_\ell^m(\hat p)
\ee
for some constants $C^{\ell m}_{i_1 \cdots i_L}$.  The values of $\ell$ that enter this sum are $\ell=L$, $L-2$, etc., down to 1 or 0 depending on whether $L$ is odd or even.  There are no values of $\ell$ greater than $L$ because $\hat p$ has $\ell=1$ and the combination of $L$ objects having $\ell=1$ can lead to angular momentum $L$ at the most.  Matching the parity $(-1)^L$ of $\hat p_{i_1} \cdots \hat p_{i_L}$ to the parity $(-1)^\ell$ of $Y_\ell^m(\hat p)$ gives the requirement that only odd or only even values of $\ell$ can contribute.  We define $\left (\hat p_{i_1} \cdots \hat p_{i_L} \right )^L_\ell$ to be the component of $\hat p_{i_1} \cdots \hat p_{i_L}$ of angular momentum $\ell$
\be \label{def.of.Cs}
\left (\hat p_{i_1} \cdots \hat p_{i_L} \right )^L_\ell \equiv \sum_{m=-\ell}^\ell C^{\ell m}_{i_1 \cdots i_L} Y_\ell^m(\hat p) ,
\ee
so that
\be
\hat p_{i_1} \cdots \hat p_{i_L} = \sum_{\ell=L}^{1 \, {\rm or} \, 0} \left (\hat p_{i_1} \cdots \hat p_{i_L} \right )^L_\ell .
\ee
It follows that any transform of the form given in \eqref{FTC} can be expressed as a linear combination of transforms like
\be \label{FT.Y}
I_{n \ell m}(\vec r\,) = \int \frac{d^3 p}{(2 \pi)^3} e^{i \vec p \cdot \vec r} p^n Y_\ell^m(\hat p).
\ee
It is convenient to express the exponential in \eqref{FT.Y} as a Rayleigh expansion: \cite{Watson22,Morse53a,Arfken01a}
\be \label{Rayleigh}
e^{i \vec p \cdot \vec r\,} = \sum_{\ell=0}^\infty i^\ell (2\ell+1) j_\ell(p r) P_\ell(\hat p \cdot \hat r)
\ee
where the $j_\ell(x)$ are spherical Bessel functions \cite{Arfken01b} and the $P_\ell(x)$ are Legendre polynomials.  We substitute \eqref{Rayleigh} into \eqref{FT.Y} and use the addition theorem of spherical harmonics \cite{Arfken01c}
\be
P_\ell(\hat p \cdot \hat r) = \frac{4 \pi}{2 \ell+1} \sum_{m=-\ell}^\ell Y_\ell^m(\hat r) Y_\ell^{m *} (\hat p)
\ee
to factor the angular dependence present in $P_\ell(\hat p \cdot \hat r)$ into parts involving the angles of $\hat p$ and $\hat r$ separately.  We integrate over the angles of $\hat p$ using orthogonality
\be
\int d \Omega_p Y_\ell^{m *} (\hat p) Y_{\ell'}^{m'} (\hat p) = \delta_{\ell \ell'} \delta_{m m'} ,
\ee
where $d \Omega_p = d \theta_p \sin \theta_p d \phi_p$ is the element of solid angle for $\hat p$,
to write the transform as
\be
I_{n \ell m}(\vec r\,) = \frac{i^\ell}{2 \pi^2} Y_\ell^m(\hat r) \int_0^\infty dp \, p^{n+2} j_\ell(p r) .
\ee
The integral
\be \label{R.integral.def}
R_{n \ell}(r) \equiv \int_0^\infty dp \, p^{n+2} j_\ell(p r)
\ee
converges for $-(\ell+3) < n < -1$ (when $n$ and $\ell$ are real, as here), and has the value $R_{n \ell}(r) = \chi_{n \ell}/r^{n+3}$ where \cite{Gradshteyn80a}
\be \label{chi.integral}
\chi_{n \ell} = 2^{n+1} \sqrt{\pi} \, \frac{\Gamma \left ( \frac{\ell+3+n}{2} \right )}{\Gamma \left ( \frac{\ell-n}{2} \right )} .
\ee
We can extend the useful range of $n$ by generalizing \eqref{R.integral.def} to
\be
R_{n \ell}(r) = \lim_{\lambda \rightarrow 0^+} \int_0^\infty dp \, e^{- \lambda p} p^{n+2} j_\ell(p r) ,
\ee
which is also given by $R_{n \ell}(r) = \chi_{n \ell}/r^{n+3}$ for all $n$ in the larger range $-(\ell+3) < n < \ell$.  When $n=\ell$ the integral contains a delta function\footnote{When $\ell$ is not an odd integer we use $(2\ell+1)!! \equiv 2^{\ell+1} \Gamma(\ell+3/2)/\Gamma(1/2)$.}: \cite{Adkins13}
\be \label{R.integral.delta}
R_{\ell \ell}(r) = \lim_{\lambda \rightarrow 0^+} \int_0^\infty dp \, e^{- \lambda p} p^{\ell+2} j_\ell(p r) = \frac{2 \pi^2 (2 \ell+1)!!}{r^\ell} \delta(\vec r\, ) .
\ee
We always integrate over the spherical angles before doing the radial integration as part of the definition of these possibly singular integrals.  (Non-spherical regularization alternatives have been considered by Hnizdo. \cite{Hnizdo11}  More general results for Fourier transforms of the form \eqref{FT.Y} have been obtained by Samko. \cite{Samko78})  All in all, we see that the initial transform \eqref{FTC} can be written as
\be
I_{n; i_1 \cdots i_L}(\vec r\, ) = \int \frac{d^3 p}{(2 \pi)^3} e^{i \vec p \cdot \vec r} \, p^n \hat p_{i_1} \cdots \hat p_{i_L} = \sum_{\ell=L}^{1 \, {\rm or} \, 0} \frac{i^\ell}{2 \pi^2} R_{n \ell}(r) \left ( \hat x_{i_1} \cdots \hat x_{i_L} \right )^L_\ell ,
\ee
where $\left ( \hat x_{i_1} \cdots \hat x_{i_L} \right )^L_\ell$ is defined in terms of $Y_\ell^m(\hat r)$ just as $\left ( \hat p_{i_1} \cdots \hat p_{i_L} \right )^L_\ell$ is in terms of $Y_\ell^m(\hat p)$.  It follows that if we can arrive at a useful expression for $\left ( \hat p_{i_1} \cdots \hat p_{i_L} \right )^L_\ell$, then we will be able to perform Fourier transforms of the form shown in \eqref{FTC} in a systematic way.


\section{Angular decomposition of $\hat p_{i_1} \cdots \hat p_{i_L}$}
\label{sec3}

Our goal in this section is to obtain an explicit and useful expression for the component $\left ( \hat p_{i_1} \cdots \hat p_{i_L} \right )^L_\ell$ of angular momentum $\ell$.  We will discuss both a constructive method most useful for low values of $L$ and a general result valid for all $L$.  Both approaches make use of the explicit solution for the constants $C^{\ell m}_{i_1 \cdots i_L}$ in \eqref{def.of.Cs}:
\be \label{soln.for.Cs}
C^{\ell m}_{i_1 \cdots i_L} = \int d \Omega_p \, Y_\ell^{m *}(\hat p) \, \hat p_{i_1} \cdots \hat p_{i_L} ,
\ee
obtained through use of the orthogonality of the spherical harmonics.  From this it is easy to see that the constants $C^{\ell m}_{i_1 \cdots i_L}$, and thus the components $\left ( \hat p_{i_1} \cdots \hat p_{i_L} \right )^L_\ell$, are completely symmetric in all indices.  The constructive method also uses the tracelessness of the maximum angular momentum component $\left ( \hat p_{i_1} \cdots \hat p_{i_L} \right )^L_L$, which follows from the tracelessness of $C^{L m}_{i_1 \cdots i_L}$, which is a consequence of the fact that an object composed of $L-2$ parts, each part of unit angular momentum, has no overlap with an object of angular momentum L:
\be \label{tracelessness}
C^{L m}_{i_1 \cdots i_L}  \delta_{i_{L-1} i_L} = \int d \Omega_p \, Y_L^{m *}(\hat p) \, \hat p_{i_1} \cdots \hat p_{i_{L-2}} = 0 .
\ee

The most convenient way to obtain $\left ( \hat p_{i_1} \cdots \hat p_{i_L} \right )^L_\ell$ for small values of $L$ is by straightforward construction.  We illustrate the constructive approach with a number of examples.  The procedure starts with the maximum angular momentum component $\left ( \hat p_{i_1} \cdots \hat p_{i_L} \right )^L_L$, which is completely symmetric and traceless.  This maximum angular momentum component can be written as $\hat p_{i_1} \cdots \hat p_{i_L}$ plus a linear combination of symmetric terms involving fewer momentum factors (but still with the same parity) added in with unknown coefficients.  The condition of tracelessness determines the coefficients.

As a first example of the constructive approach, consider the case $L=3$.  The maximum angular momentum component is
\be
\left ( \hat p_i \hat p_j \hat p_k \right )^3_3 = \hat p_i \hat p_j \hat p_k - \frac{1}{5} \left (\hat p_i \delta_{j k} + \hat p_j \delta_{k i} + \hat p_k \delta_{i j} \right ) ,
\ee
where the $-1/5$ coefficient was determined by the tracelessness condition.  The other component, $\left ( \hat p_i \hat p_j \hat p_k \right )^3_1$, is the difference $\hat p_i \hat p_j \hat p_k - \left ( \hat p_i \hat p_j \hat p_k \right )^3_3$:
\be
\left ( \hat p_i \hat p_j \hat p_k \right )^3_1 = \frac{1}{5} \left (\hat p_i \delta_{j k} + \hat p_j \delta_{k i} + \hat p_k \delta_{i j} \right ) .
\ee
It is clear that $\left ( \hat p_i \hat p_j \hat p_k \right )^3_1$ has $\ell=1$ because each of its terms is linear in $\hat p$.

As a second example of explicit construction, we consider the term with $L=4$.  The term with maximal angular momentum is
\be \label{eg.4.4}
\left ( \hat p_{i_1} \hat p_{i_2} \hat p_{i_3} \hat p_{i_4} \right )^4_4 = \hat p_{i_1} \hat p_{i_2} \hat p_{i_3} \hat p_{i_4} - \frac{1}{7} \left ( \hat p_{i_1} \hat p_{i_2} \delta_{i_3 i_4} + {\rm perms} \right )_{\rm 6 \; terms} + \frac{1}{35} \left ( \delta_{i_1 i_2} \delta_{i_3 i_4} + {\rm perms} \right )_{\rm 3 \; terms} ,
\ee
where the coefficients $-1/7$ and $1/35$ were obtained by applying the tracelessness condition.  We are only writing one representative permutation of indices--the others are represented by ``$+ \,{\rm perms}$'' and an indication of how many independent permutations in all there are.  We identify the $\ell=2$ component $\left ( \hat p_{i_1} \hat p_{i_2} \hat p_{i_3} \hat p_{i_4} \right )^4_2$ by isolating the term in the difference $\hat p_{i_1} \hat p_{i_2} \hat p_{i_3} \hat p_{i_4} - \left ( \hat p_{i_1} \hat p_{i_2} \hat p_{i_3} \hat p_{i_4} \right )^4_4$ that is quadratic in $\hat p$ and subtracting the appropriate momentum-independent terms so that each part of $\left ( \hat p_{i_1} \hat p_{i_2} \hat p_{i_3} \hat p_{i_4} \right )^4_2$ has $\ell=2$:
\be \label{eg.4.2}
\left ( \hat p_{i_1} \hat p_{i_2} \hat p_{i_3} \hat p_{i_4} \right )^4_2 = \frac{1}{7} \left ( \left ( \hat p_{i_1} \hat p_{i_2} \right )^2_2 \delta_{i_3 i_4} + {\rm perms} \right )_{\rm 6 \; terms} .
\ee
The $\ell=0$ component is the remainder:
\be \label{eg.4.0}
\left ( \hat p_{i_1} \hat p_{i_2} \hat p_{i_3} \hat p_{i_4} \right )^4_0 = \frac{1}{15} \left ( \delta_{i_1 i_2} \delta_{i_3 i_4} + {\rm perms} \right )_{\rm 3 \; terms} .
\ee

We have also constructed the $L=5$ decomposition and just give the results:
\bse
\bearray
\left ( \hat p_{i_1} \hat p_{i_2} \hat p_{i_3} \hat p_{i_4} \hat p_{i_5} \right )^5_5 &=& \hat p_{i_1} \hat p_{i_2} \hat p_{i_3} \hat p_{i_4} \hat p_{i_5} - \frac{1}{9} \left ( \hat p_{i_1} \hat p_{i_2} \hat p_{i_3} \delta_{i_4 i_5} + {\rm perms} \right )_{\rm{10 \; terms}} + \frac{1}{63} \left ( \hat p_{i_1} \delta_{i_2 i_3} \delta_{i_4 i_5} + {\rm perms} \right )_{\rm{15 \; terms}} , \label{eg.5.5} \\
\left ( \hat p_{i_1} \hat p_{i_2} \hat p_{i_3} \hat p_{i_4} \hat p_{i_5} \right )^5_3 &=& \frac{1}{9} \big ( \left ( \hat p_{i_1} \hat p_{i_2} \hat p_{i_3} \right )^3_3 \delta_{i_4 i_5} + {\rm perms} \big )_{\rm{10 \; terms}} , \label{eg.5.3} \\
\left ( \hat p_{i_1} \hat p_{i_2} \hat p_{i_3} \hat p_{i_4} \hat p_{i_5} \right )^5_1 &=& \frac{1}{35} \big ( \hat p_{i_1} \delta_{i_2 i_3} \delta_{i_4 i_5} + {\rm perms} \big )_{\rm{15 \; terms}} . \label{eg.5.1}
\eearray
\ese

The basis of our general construction of $\left ( \hat p_{i_1} \cdots \hat p_{i_L} \right )^L_\ell$ is an inductive argument using a recursion relation giving a component with angular momentum $\ell$ in terms of components with lower values of $\ell$.  We will propose a general expression for $\left ( \hat p_{i_1} \cdots \hat p_{i_L} \right )^L_\ell$ and show that it satisfies both the recursion relation and the appropriate initial values.

We can obtain a useful expression for $\left ( \hat p_{i_1} \cdots \hat p_{i_L} \right )^L_\ell$ by using \eqref{soln.for.Cs} in \eqref{def.of.Cs} along with the addition theorem for spherical harmonics:
\bearray \label{integral.soln}
\left ( \hat p_{i_1} \cdots \hat p_{i_L} \right )^L_\ell &=& \int d \Omega_{p'} \, \hat p'_{i_1} \cdots \hat p'_{i_L} \sum_{m=-\ell}^\ell Y_\ell^{m *}(\hat p') Y_\ell^m(\hat p) \cr
&=&  (2\ell+1) \int \frac{d \Omega_{p'}}{4 \pi} \, \hat p'_{i_1} \cdots \hat p'_{i_L} \, P_\ell(\hat p' \cdot \hat p) .
\eearray
It can be seen from this expression both that $\left ( \hat p_{i_1} \cdots \hat p_{i_L} \right )^L_\ell=0$ for $\ell>L$, and that $\left ( \hat p_{i_1} \cdots \hat p_{i_L} \right )^L_\ell=0$ unless $L-\ell$ is even (by use of a parity argument).  For any particular value of $\ell$ we could write out $P_\ell(\hat p' \cdot \hat p)$ as a polynomial of order $\ell$ and perform the angular integral using \cite{Bowen94}
\be
\int \frac{d \Omega}{4 \pi} \, \hat x_{i_1} \cdots \hat x_{i_N} = \frac{\delta_{N,{\rm even}}}{(N+1)!!} \left ( \delta_{i_1 i_2} \cdots \delta_{i_{N-1} i_N} + {\rm perms} \right )_{(N-1)!! \; {\rm terms}} \, .
\ee
It is easy to perform the integral in \eqref{integral.soln} for $\ell=0$ and $\ell=1$ where $P_0(\hat p' \cdot \hat p)=1$ and $P_1(\hat p' \cdot \hat p)=\hat p' \cdot \hat p = \hat p'_j \hat p_j$ (with an implied sum over $j$ from 1 to 3).  One finds that
\bse \label{gen.form.init}
\bearray
\left ( \hat p_{i_1} \cdots \hat p_{i_L} \right )^L_0 &=& \int \frac{d \Omega_{p'}}{4 \pi} \, \hat p'_{i_1} \cdots \hat p'_{i_L} \cr
& =& \frac{\delta_{L,{\rm even}}}{(L+1)!!} \left ( \delta_{i_1 i_2} \cdots \delta_{i_{L-1} i_L} + {\rm perms} \right )_{(L-1)!! \; {\rm terms}}  ,  \\
\left ( \hat p_{i_1} \cdots \hat p_{i_L} \right )^L_1 &=& 3 \int \frac{d \Omega_{p'}}{4 \pi} \, \hat p'_{i_1} \cdots \hat p'_{i_L} \hat p'_j \, \hat p_j  = 3 \left ( \hat p_{i_1} \cdots \hat p_{i_L} \hat p_j \right )^{L+1}_0 \, \hat p_j \nonumber \\[1pt]
& =& \frac{3 \delta_{L,{\rm odd}}}{(L+2)!!} \Bigl ( \hat p_{i_1} \left ( \delta_{i_2 i_3} \cdots \delta_{i_{L-1} i_L} + {\rm perms} \right )_{(L-2)!! \; {\rm terms}}  + {\rm perms} \Bigr )_{L \; {\rm terms}} \, .
\eearray
\ese
These results will serve as initial values for the inductive argument.  The recursion relation for $\left ( \hat p_{i_1} \cdots \hat p_{i_L} \right )^L_\ell$ is
\be \label{recursion}
\left ( \hat p_{i_1} \cdots \hat p_{i_L} \right )^L_\ell = \frac{2\ell+1}{\ell} \left \{ \left ( \hat p_{i_1} \cdots \hat p_{i_L} \hat p_j \right )^{L+1}_{\ell-1} \hat p_j - \frac{\ell-1}{2 \ell-3} \left ( \hat p_{i_1} \cdots \hat p_{i_L} \right )^L_{\ell-2} \right \} ,
\ee
which follows from \eqref{integral.soln} and the recursion relation for Legendre polynomials
\be
\ell P_\ell(x) = (2\ell-1) x P_{\ell-1}(x) - (\ell-1) P_{\ell-2}(x) .
\ee

In order to write down a general expression for $\left ( \hat p_{i_1} \cdots \hat p_{i_L} \right )^L_\ell$ and perform the inductive proof of its correctness it will be useful to introduce a little notation.  First, we will use the usual summation symbol to represent a `sum over permutations' instead of the `$+$ perms' notation used up until now.  Specifically, we will write
\be \label{def.perm.sum}
\sum_{(L-1)!!} \Big ( \delta_{i_1 i_2} \cdots \delta_{i_{L-1} i_L} \Big )
\ee
for the sum over the $(L-1)!!$ independent permutations of the indices of the quantity in parentheses.  This sum over permutations doesn't include ones that trivially equal one another, which is why there are only $(L-1)!!$ permutations in \eqref{def.perm.sum}, but would be $L!$ permutations in a sum over permutations of $A^1_{i_1} \cdots A^L_{i_L}$, but only one in a sum over permutations of $S_{i_1 \cdots i_L}$ if $S_{i_1 \cdots i_L}$ is completely symmetric.  Also, we define the new symbol
\be
X^{L, \ell}_{i_1 \cdots i_L} \equiv \sum_{\genfrac{(}{)}{0pt}{}{L}{\ell}} \Bigl ( \hat p_{i_1} \cdots \hat p_{i_\ell} \sum_{(L-\ell-1)!!} \Big ( \delta_{i_{\ell+1} i_{\ell+2}} \cdots \delta_{i_{L-1} i_L} \Big ) \Big )
\ee
to represent the completely symmetric object with $L$ indices formed of $\ell$ momentum unit vectors and $(L-\ell)/2$ Kronecker deltas.  The $\genfrac{(}{)}{0pt}{}{L}{\ell} $ notation represents the combinatoric factor $\frac{L!}{\ell! (L-\ell)!}$ for the number of ways to pick $\ell$ indices out of a collection of $L$ indices. We can represent $X^{L, \ell}_{i_1 \cdots i_L}$ slightly more compactly as
\be
X^{L, \ell}_{i_1 \cdots i_L} = \sum_{\genfrac{(}{)}{0pt}{}{L}{\ell} (L-\ell-1)!!} \Bigl ( \hat p_{i_1} \cdots \hat p_{i_\ell} \delta_{i_{\ell+1} i_{\ell+2}} \cdots \delta_{i_{L-1} i_L} \Big ) .
\ee
We note that the $X^{L, \ell}_{i_1 \cdots i_L}$ symbol only makes sense when $L$ and $\ell$ are either both even or both odd--we define it to be zero otherwise.  We also define $X^{L, \ell}_{i_1 \cdots i_L}$ to be zero if $L$ or $\ell$ is negative or if $\ell$ is greater than $L$.  With the new notation we can write the results of \eqref{gen.form.init} as
\bse
\bearray
\left ( \hat p_{i_1} \cdots \hat p_{i_L} \right )^L_0 &=&  \frac{1}{(L+1)!!} X^{L,0}_{i_1 \cdots i_L} \; , \label{gen.form.ell0} \\
\left ( \hat p_{i_1} \cdots \hat p_{i_L} \right )^L_1 &=& \frac{3}{(L+2)!!} X^{L,1}_{i_1 \cdots i_L} . \label{gen.form.ell1}
\eearray
\ese

Identities among $X^{L, \ell}_{i_1 \cdots i_L}$ quantities can often be found by simple counting.  For instance, consider the following identity for the symmetrized product of two $X$s:
\be \label{product.identity}
\sum_{\genfrac{(}{)}{0pt}{}{L+N}{L}} \Bigl ( X^{L, \ell}_{i_1 \cdots i_L} X^{N, n}_{i_{L+1} \cdots i_{L+N}} \Bigr ) = \kappa X^{L+N, \ell+n}_{i_1 \cdots i_{L+N}} .
\ee
Both sides of \eqref{product.identity} involve the same set of $L+N$ indices, both are symmetric in all indices, and both have exactly $\ell+n$ momentum unit vectors in each term, so the two sides of \eqref{product.identity} are proportional.  Since all terms enter with the same sign and there are no cancellations, the constant of proportionality $\kappa$ can be found simply by counting the total number of terms on each side.  On the left there are $\left ( {{L+N} \atop L} \right ) \left ( {L \atop \ell} \right ) (L-\ell-1)!! \left ( {N \atop n} \right ) (N-n-1)!!$ terms, while on the right there are $\left ( {L+N} \atop {\ell+n} \right ) (L+N-\ell-n-1)!!$ terms.  The constant $\kappa$ is the ratio: 
\be \label{kappa}
\kappa = \left ( {L+N-\ell-n} \atop {L-\ell} \right ) \left ( {\ell+n} \atop \ell \right ) \frac{(L-\ell-1)!! (N-n-1)!!}{(L+N-\ell-n-1)!!} .
\ee

Two additional identities that will be useful to us involve the contraction $X^{L, \ell}_{i_1 \cdots i_L} \hat p_{i_L}$ of an $X$ with $\hat p$ and the contraction $X^{L, \ell}_{i_1 \cdots i_L} \delta_{i_{L-1} i_L}$ of two indices of an $X$.  For the first identity, we note that 
\bearray \label{Xp.init}
X^{L, \ell}_{i_1 \cdots i_L} &=& \sum_{\left ( {L-1} \atop \ell \right )} \Bigl ( \hat p_{i_1} \cdots \hat p_{i_\ell} \sum_{(L-\ell-1)!!} \Bigl ( \delta_{i_{\ell+1} i_{\ell+2}} \cdots \delta_{i_{L-1} i_L} \Bigr ) \Bigr ) \cr
& +& \sum_{\left ( {L-1} \atop { \ell-1} \right )} \Bigl ( \hat p_{i_1} \cdots \hat p_{i_{\ell-1}} \hat p_{i_L} \sum_{(L-\ell-1)!!} \Bigl ( \delta_{i_{\ell} i_{\ell+1}} \cdots \delta_{i_{L-2} i_{L-1}} \Bigr ) \Bigr ) ,
\eearray
where in the first term it is understood that index $i_L$ is definitely on a $\delta$, while in the second term index $i_L$ is attached to a $\hat p$.  Contraction with $\hat p_{i_L}$ then gives two corresponding terms: $\hat p_{i_L}$ times the first term of \eqref{Xp.init} has $L-1$ indices and $\ell+1$ factors of $\hat p$, while $\hat p_{i_L}$ times the second term of \eqref{Xp.init} has $L-1$ indices and only $\ell-1$ powers of $\hat p$, so that $X^{L, \ell}_{i_1 \cdots i_L} \hat p_{i_L} = \alpha X^{L-1, \ell+1}_{i_1 \cdots i_{L-1}} + \beta X^{L-1, \ell-1}_{i_1 \cdots i_{L-1}}$ for some constants $\alpha$ and $\beta$.  Counting terms allows us to identity the constants to be $\alpha=\ell+1$ and $\beta=1$, so that
\be \label{Xp}
X^{L, \ell}_{i_1 \cdots i_L} \hat p_{i_L} =  (\ell+1) X^{L-1, \ell+1}_{i_1 \cdots i_{L-1}} + X^{L-1, \ell-1}_{i_1 \cdots i_{L-1}} .
\ee
For the second identity we write $X^{L, \ell}_{i_1 \cdots i_L}$ as
\bearray \label{Xtr.init}
X^{L, \ell}_{i_1 \cdots i_L} &=& \sum_{\left ( {L-2} \atop \ell \right )} \Biggl ( \hat p_{i_1} \cdots \hat p_{i_\ell} \Bigl \{ \sum_{(L-\ell-3)!!} \Bigl ( \delta_{i_{\ell+1} i_{\ell+2}} \cdots \delta_{i_{L-1} i_L} \Bigr ) \cr
& \quad \quad \quad \quad \quad +& \sum_{(L-\ell-2)(L-\ell-3)!!} \Bigl ( \delta_{i_{\ell+1} i_{\ell+2}} \cdots \delta_{i_{L-3} i_{L-1}} \delta_{i_{L-2} i_L} \Bigr ) \Bigr \} \Biggr )\cr
& \quad +& \sum_{\left ( {L-2} \atop {\ell-1} \right )} \Bigl ( \hat p_{i_1} \cdots \hat p_{i_{\ell-1}} \hat p_{i_{L-1}} \sum_{(L-\ell-1)!!} \Bigl ( \delta_{i_{\ell} i_{\ell+1}} \cdots \delta_{i_{L-2} i_{L}} \Bigr ) \Bigr ) \cr
& \quad +& \sum_{\left ( {L-2} \atop {\ell-1} \right )} \Bigl ( \hat p_{i_1} \cdots \hat p_{i_{\ell-1}} \hat p_{i_{L}} \sum_{(L-\ell-1)!!} \Bigl ( \delta_{i_{\ell} i_{\ell+1}} \cdots \delta_{i_{L-2} i_{L-1}} \Bigr ) \Bigr ) \cr
& \quad +& \sum_{\left ( {L-2} \atop {\ell-2} \right )} \Bigl ( \hat p_{i_1} \cdots \hat p_{i_{\ell-2}} \hat p_{i_{L-1}} \hat p_{i_L} \sum_{(L-\ell-1)!!} \Bigl ( \delta_{i_{\ell-1} i_{\ell}} \cdots \delta_{i_{L-3} i_{{L-2}}} \Bigr ) \Bigr ) ,
\eearray
where it is understood that in the first two terms the indices $i_{L-1}$ and $i_L$ are definitely on $\delta$s, on the same $\delta$ in the first and on different $\delta$s in the second; in the third term $i_{L-1}$ is on a $\hat p$ while $i_L$ is on a $\delta$; in the fourth $i_L$ is on a $\hat p$ and $i_{L-1}$ on a $\delta$; and in the last term both $i_{L-1}$ and $i_L$ are on $\hat p$s.  Contraction of $i_{L-1}$ with $i_L$ gives rise to two structures: $X^{L, \ell}_{i_1 \cdots i_L} \delta_{i_{L-1} i_L} = \rho X^{L-2, \ell}_{i_1 \cdots i_{L-2} }+ \sigma X^{L-2, \ell-2}_{i_1 \cdots i_{L-2}}$, with the first four terms of \eqref{Xtr.init} contributing to $\rho$ and only the last contributing to $\sigma$.  Again, we use term counting to identify values for $\rho$ and $\sigma$, finding $\rho=L+\ell+1$ and $\sigma=1$.  (The four contributions to $\rho$ are, in order, $3$, $L-\ell-2$, $\ell$ and $\ell$.)  The final form of the contraction identity is
\be \label{Xtr}
X^{L, \ell}_{i_1 \cdots i_L} \delta_{i_{L-1} i_L} =  (L+\ell+1) X^{L-2, \ell}_{i_1 \cdots i_{L-2}} + X^{L-2, \ell-2}_{i_1 \cdots i_{L-2}} .
\ee

We propose the following form for the general decomposition formula
\be \label{decomp}
\left ( \hat p_{i_1} \cdots \hat p_{i_L} \right )^L_\ell = \frac{(2\ell+1)(L-\ell-1)!!}{(L-\ell)! (L+\ell+1)!!} \sum_{n=\ell}^{1 \, {\rm or} \, 0} (-1)^\frac{\ell-n}{2} \frac{(L-n)! (\ell+n-1)!! (\ell-n-1)!!}{(\ell-n)! (L-n-1)!!} X^{L,n}_{i_1 \cdots i_L} ,
\ee
where the sum is over $n=\ell$, $\ell-2$, etc., ending with 1 or 0 depending on whether $\ell$ is odd or even.  We proceed to verify the correctness of this formula by: ($i$), showing that it is consistent with the initial values of \eqref{gen.form.ell0}, \eqref{gen.form.ell1}; and ($ii$), verifying that it satisfies the recursion relation \eqref{recursion}.  Verification of consistency with the initial values is immediate by substituting $\ell=0$ and $\ell=1$ into \eqref{decomp} and noting that in each case there is only one term in the sum and that it agrees with \eqref{gen.form.ell0}, \eqref{gen.form.ell1}.  For step ($ii$), verification that \eqref{decomp} satisfies the recursion relation \eqref{recursion}, we substitute \eqref{decomp} into the right hand side of \eqref{recursion} and obtain two terms.  The first of these is
\small \be \label{term1}
\frac{2\ell+1}{\ell} \frac{(2\ell-1)(L-\ell+1)!!}{(L-\ell+2)! (L+\ell+1)!!} \sum_{n'=\ell-1}^{0 \, {\rm or} \, 1} (-1)^\frac{\ell-n'-1}{2} \frac{(L-n'+1)! (\ell+n'-2)!! (\ell-n'-2)!!}{(\ell-n'-1)! (L-n')!!} \left ( (n'+1) \, X^{L,n'+1}_{i_1 \cdots i_L} + X^{L, n'-1}_{i_1 \cdots i_L} \right )
\ee \normalsize
and the second is
\small \be \label{term2}
- \frac{2\ell+1}{\ell} \frac{\ell-1}{2\ell-3} \frac{(2\ell-3)(L-\ell+1)!!}{(L-\ell+2)! (L+\ell-1)!!} \sum_{n=\ell-2}^{1 \, {\rm or} \, 0} (-1)^\frac{\ell-n-2}{2} \frac{(L-n)! (\ell+n-3)!! (\ell-n-3)!!}{(\ell-n-2)! (L-n-1)!!}  \, X^{L,n}_{i_1 \cdots i_L} \, .
\ee \normalsize
We shift the summation index in the first part of \eqref{term1} according to $n' \rightarrow n-1$ and in the second part by $n' \rightarrow n+1$, so that all terms are proportional to $X^{L,n}_{i_1 \cdots i_L}$.  We add the two parts of \eqref{term1} to \eqref{term2} and, after some algebraic simplification, find that the sum is equal to \eqref{decomp}.  Thus \eqref{decomp} satisfies the recursion relation and by induction is correct for all values of $\ell$. \footnote{An expression consistent with \eqref{decomp} is given in \cite{Bezubik04,Bezubik06} but more general because not restricted to three dimensions of space.  The consistency of \eqref{decomp} with Theorem 1 of \cite{Bezubik04,Bezubik06} is established through use of the identity $\Delta^n p_{i_1} \cdots p_{i_L} = 2^n n! p^{L-2n} X^{L,L-2n}_{i_1 \cdots i_L}$ where $\Delta$ is the Laplacian.}

It is useful to find an expression for the component of $\hat p_{i_1} \cdots \hat p_{i_L}$ having maximal angular momentum.  With $\ell \rightarrow L$ in the general decomposition formula \eqref{decomp} we find that
\be \label{max.ell}
\left ( \hat p_{i_1} \cdots \hat p_{i_L} \right )^L_L = \sum_{n=L}^{1 \, {\rm or} \, 0} (-1)^\frac{L-n}{2} \frac{(L+n-1)!!}{(2L-1)!!} X^{L,n}_{i_1 \cdots i_L} .
\ee
This expression is traceless on all pairs of indices as required by \eqref{tracelessness} and as can be confirmed by applying the trace identity \eqref{Xtr} to \eqref{max.ell}.

Using \eqref{max.ell}, the product identity \eqref{product.identity}, and expression \eqref{gen.form.ell0} for $\left ( \hat p_{i_1} \cdots \hat p_{i_L} \right )^L_0$, it is easy to see that $\left ( \hat p_{i_1} \cdots \hat p_{i_L} \right )^L_\ell$ can be written in an alternative, and illuminating, form:
\be \label{decomp.ell}
\left ( \hat p_{i_1} \cdots \hat p_{i_L} \right )^L_\ell = \frac{(2 \ell+1)!! (L-\ell+1)!!}{(L+\ell+1)!!} \sum_{\begin{pmatrix} L \\ \ell \end{pmatrix}} \left ( \hat p_{i_1} \cdots \hat p_{i_\ell} \right )^\ell_\ell \left ( \hat p_{i_{\ell+1}} \cdots \hat p_{i_L} \right )^{L-\ell}_0 \, .
\ee
This form displays clearly the fact that every sub-part of $\left ( \hat p_{i_1} \cdots \hat p_{i_L} \right )^L_\ell$ has angular momentum $\ell$ among some subset of momenta unit vectors and angular momentum zero among the rest.  This behavior is illustrated by the examples shown in \eqref{eg.4.2}, \eqref{eg.5.3}, and \eqref{eg.5.1} (since $(\hat p_i)^1_1=\hat p_i$).  Expressions \eqref{max.ell} for $\left ( \hat p_{i_1} \cdots \hat p_{i_L} \right )^L_L$ and \eqref{decomp.ell} for $\left ( \hat p_{i_1} \cdots \hat p_{i_L} \right )^L_\ell$ are the main results of this paper.


\section{Applications}
\label{sec4}

As discussed in \cite{Adkins13} and \ref{sec2}, a useful class of three-dimensional Fourier transforms (that of functions like $p^n \hat p_{i_1} \cdots \hat p_{i_L}$) can be conveniently done after separation of the various angular momenta in the tensor product.  With a Fourier transform pair $\Phi(\vec p\, )$ and $\Psi(\vec r\, )$ defined through
\bse \label{FT.def.vec}
\bearray
\Psi(\vec r\,) &=& \int \frac{d^3 p}{(2 \pi)^3} \, e^{i \vec p \cdot \vec r} \Phi(\vec p\,) \, , \\
\Phi(\vec p\,) &=& \int d^3 r \, e^{-i \vec p \cdot \vec r} \Psi(\vec r\,) \, ,
\eearray
\ese
it is generally true that the angular momenta of $\Phi(\vec p\, )$ and $\Psi(\vec r\, )$ are the same.  It follows that the transform pairs can be represented by
\be
\Phi(\vec p\, )=\phi(p) Y_\ell^m(\hat p) \iff \Psi(\vec r\, )=\psi(r) Y_\ell^m(\hat r)
\ee
or
\be
\Phi(\vec p\, )=\phi(p) \left ( \hat p_{i_1} \cdots \hat p_{i_L} \right )^L_\ell \iff \Psi(\vec r\, )=\psi(r) \left ( \hat x_{i_1} \cdots \hat x_{i_L} \right )^L_\ell
\ee
where the radial functions $\phi(p)$ and $\psi(r)$ are related by
\bse
\bearray 
\psi(r) &=& \frac{i^\ell}{2 \pi^2} \int_0^\infty dp \, p^2 j_\ell (p r) \phi(p) \, , \label{scalara}  \\
\phi(p) &=& 4 \pi (-i)^\ell \int_0^\infty dr \, r^2 j_\ell (p r) \psi(r) \label{scalarb} \, .
\eearray 
\ese
In the case that $\phi(p)=p^n$ the transform is $\psi(r)$ where
\be \label{transformpair}
\psi(r) =
\begin{cases}
\frac{i^\ell}{2 \pi^2} \frac{\chi_{n \ell}}{r^{n+3}}  &-(\ell+3)<n<\ell \, ,\\[0.5em]
\frac{i^\ell (2\ell+1)!!}{r^\ell} \delta(\vec r\,) & \quad \quad \quad n=\ell \, .
\end{cases}
\ee

Since the transform pair \eqref{scalara}, \eqref{scalarb} is essentially the Hankel transform, \cite{Oberhettinger72,Davies78} many additional $\phi$, $\psi$ pairs are also available, for example those that relate momentum space and coordinate space versions of the Coulomb wave functions. \cite{Podolsky29}  Results for the examples given in \ref{introduction} are immediate consequences of angular decomposition and the transforms contained in \eqref{transformpair}.

An interesting use of Fourier transforms of the type considered here is to find unusual differential identities.  Consider the Fourier transform of $f(p) \left ( p_{i_1} \cdots p_{i_L} \right )^L_\ell$.  The momentum vectors can be converted into derivatives when acting on the exponential in the  Fourier transform, leading to
\be \label{FT.derivs}
\int \frac{d^3 p}{(2 \pi)^3} \, e^{i \vec p \cdot \vec r} f(p) \left ( p_{i_1} \cdots p_{i_L} \right )^L_\ell = (-i)^L \left ( \partial_{i_1} \cdots \partial_{i_L} \right )^L_\ell \int \frac{d^3 p}{(2 \pi)^3} \, e^{i \vec p \cdot \vec r} f(p).
\ee
On the other hand, the transform can be evaluated explicitly using \eqref{transformpair}.  Comparison leads to a differential identity.  For example, comparison of the two approaches for the transform of $\frac{1}{p^2} \left ( p_{i_1} \cdots p_{i_k} \right )^k_k$ leads to the identity
\be \label{onebyr.identity}
\left ( \partial_{i_1} \cdots \partial_{i_k} \right )^k_k \frac{1}{r} = \frac{(-1)^k (2k-1)!!}{r^{k+1}} \left ( \hat x_{i_1} \cdots \hat x_{i_k} \right )^k_k .
\ee
(The same identity expressed in terms of spherical harmonics has been given by Rowe. \cite{Rowe78}  General derivatives of inverse powers of $r$ have been worked out by Estrada and Kanwal from a distribution point of view. \cite{Estrada85})  For $k=2$, and with use of the Poisson equation $\partial^2 \frac{1}{r} = -4 \pi \delta(\vec r\, )$, one obtains the familiar identity \cite{Frahm83}
\be
\partial_i \partial_j \frac{1}{r} = -\frac{4 \pi}{3} \delta_{i j} \delta(\vec r\, ) + \frac{3}{r^3} \left ( \hat x_i \hat x_j - \frac{1}{3} \delta_{i j} \right ) ,
\ee
as also follows from \eqref{FTell2} of Sec.~\ref{introduction}.  A second interesting identity of this class follows from consideration of the transform of $\left ( p_{i_1} \cdots p_{i_k} \right )^k_k$:
\be
\left ( \partial_{i_1} \cdots \partial_{i_k} \right )^k_k \delta(\vec r\,) = \frac{(-1)^k (2k+1)!!}{r^k} \left ( \hat x_{i_1} \cdots \hat x_{i_k} \right )^k_k \delta(\vec r\,) .
\ee
Other differential identities can be obtained as easily.

\begin{acknowledgments}
This material is based upon work supported by the National Science Foundation through Grant No. PHY-1404268.
\end{acknowledgments}

\break



\end{document}